\documentclass[aps,twocolumn,floatfix,superscriptaddress]{revtex4}

\usepackage[ansinew]{inputenc}
\usepackage[american]{babel}
\usepackage[T1]{fontenc}
\usepackage{graphicx}

\renewcommand{\deg}{$^\circ$}

\begin{document}

\title{Spin injection from perpendicular magnetized ferromagnetic $\delta$-MnGa into (Al,Ga)As heterostructures}

\author{C. Adelmann}
\affiliation{Department of Chemical Engineering and Materials Science}

\author{X. Lou}
\author{H.-S. Chiang}
\affiliation{School of Physics and Astronomy, University of Minnesota, Minneapolis, Minnesota 5545-0132, USA}

\author{J.L. Hilton}
\author{B.D. Schultz}
\author{S. McKernan}
\affiliation{Department of Chemical Engineering and Materials Science}

\author{P.A. Crowell}
\affiliation{School of Physics and Astronomy, University of Minnesota, Minneapolis, Minnesota 5545-0132, USA}

\author{C.J. Palmstr\o{}m}
\affiliation{Department of Chemical Engineering and Materials Science}

\begin{abstract}

Electrical spin injection from ferromagnetic $\delta$-MnGa into an (Al,Ga)As \emph{p-i-n} light emitting diode (LED) is demonstrated. The $\delta$-MnGa layers show strong perpendicular magnetocrystalline anisotropy, enabling detection of spin injection at remanence without an applied magnetic field. The bias and temperature dependence of the spin injection are found to be qualitatively similar to Fe-based spin LED devices. A Hanle effect is observed and demonstrates complete depolarization of spins in the semiconductor in a transverse magnetic field.

\end{abstract}

\maketitle

The electrical injection of non-equilibrium spin polarization from ferromagnetic metals into semiconductors is a central requirement in the field of spintronics \cite{Zutic}. Recently, spin injection from ferromagnetic metals such as Fe \cite{Zhu,Hanbicki,Adelmann,Crooker,Lou}, Fe$_x$Co$_{1-x}$ \cite{Motsnyi, Jiang}, Fe$_3$Si \cite{Kawa}, Co$_2$MnGe \cite{Dong}, and MnAs \cite{Ramsteiner} into GaAs has been demonstrated. In these experiments, the spin-injecting contacts are thin metal films with thicknesses of the order of a few nm, which show in-plane magnetization due to shape anisotropy. However, the spin-dependent optical recombination in surface-emitting devices is sensitive to the out-of-plane spin polarization \cite{DyakonovOO}, and large magnetic fields of the order of 20\,kOe are usually applied to rotate the magnetization out-of-plane. This is a particular burden for the operation of emerging magnetophotonic devices such as the spin light-emitting diode (spin LED) \cite{Zhu,Hanbicki,Adelmann,Motsnyi,Jiang,Kawa,Dong,Ramsteiner} and the spin vertical-cavity surface-emitting laser (spin VCSEL) \cite{Rudolph,Holub}, which are of great interest for applications in spintronics, including optical switches and modulators.	

To allow more flexibility in the design of spintronic devices, it would be desirable to control the direction of the spin polarization, including out-of-plane orientation, without the need for large applied magnetic fields. The practical operation of an electrically-pumped surface-light-emitting spintronic device would require a ferromagnetic metallic contact with a perpendicular magnetic anisotropy large enough to overcome the shape anisotropy as well as a Curie temperature above room temperature. $\delta$-Mn$_x$Ga$_{1-x}$ ($0.55<x<0.60$) fulfills these requirements, and it has been grown epitaxially on GaAs (001) \cite{Krishnan,Tanaka,Hilton1}. In this letter, we demonstrate the injection of an out-of-plane spin polarization from $\delta$-Mn$_{0.58}$Ga$_{0.42}$ (henceforth ``$\delta$-MnGa'') into (Al,Ga)As heterostructures in remanence. The spin polarization at remanence is greater than 4\% at low temperatures \cite{Gerhardt}. The polarization is completely destroyed by a small transverse magnetic field, demonstrating the existence of a Hanle effect.

The $\delta$-MnGa films were grown by molecular-beam epitaxy on (001) GaAs substrates. For spin injection, the ferromagnetic metal/semiconductor interface is believed to be critical. Although $\delta$-MnGa is thermodynamically stable on GaAs, Mn is not \cite{Hilton1}, reacting with GaAs to form $\delta$-MnGa and Mn$_2$As \cite{Hilton1}. Since Mn$_2$As is an antiferromagnet, its presence at the interface may be expected to be detrimental to spin injection. Hence, in order to minimize Mn/GaAs interfacial reactions, an initial template layer growth of $\delta$-MnGa was performed at low temperature, following the procedure introduced by Tanaka \emph{et al.} \cite{Tanaka}. In order to mimic the atomic stacking sequence of MnGa in the [001] direction, alternating layers of Mn (three) and Ga (two) were deposited at 40\,\deg C, forming an amorphous template layer on the substrate. This template layer was annealed at 225\,\deg C for 15 minutes, causing the film to crystallize and nucleate on the GaAs surface by solid-phase epitaxy, in order to promote the formation of a thermodynamically stable epitaxial interlayer. The growth of $\delta$-MnGa is continued by Mn and Ga co-deposition at 175\,\deg C. Finally, the films were annealed in situ at 325\,\deg C for two minutes to improve the overall ordering and the surface smoothness. 

The epitaxial quality of the $\delta$-MnGa/GaAs heterostructure was determined by X-ray diffraction and cross-sectional high-resolution transmission electron microscopy (HRTEM). The X-ray diffraction pattern of a 10\,nm thick $\delta$-MnGa film in Fig.~1(a) shows peaks corresponding to the (001) and (002) planes and demonstrates epitaxial growth of the desired tetragonal $\delta$-MnGa phase, with $\delta$-MnGa (001) $\parallel$ GaAs (001) and an out-of-plane lattice parameter of $c = 0.368$\,nm. The cross-sectional HRTEM micrograph of a 10\,nm $\delta$-MnGa/GaAs (001) heterostructure in Fig.~2(b) shows that the film is epitaxial. A close examination, however, suggests that an interfacial layer with a thickness of $\approx 1$\,nm exists between $\delta$-MnGa and GaAs. The interfacial layer may stem from the template growth, an interpretation that is supported by scanning tunneling microscopy results \cite{Hilton2}. However, more studies are necessary to elucidate the detailed interfacial structure of $\delta$-MnGa/GaAs heterostructures.

\begin{figure}[tb]
\includegraphics[width=8cm,clip]{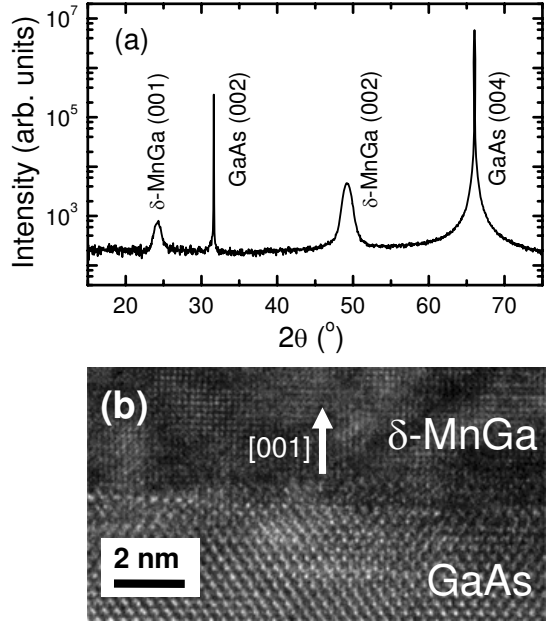}
\caption{\small \textbf{(a)} X-Ray diffraction pattern of a 10\,nm thick $\delta$-MnGa film on GaAs (001), indicating epitaxial growth with a (001) orientation. \textbf{(b)} Cross-sectional high-resolution transmission electron micrograph of the interface of a 10\,nm thick $\delta$-MnGa film on GaAs (001).}
\end{figure}

Spin injection from $\delta$-MnGa into (Al,Ga)As heterostructures was assessed using polarized electroluminescence (EL). 7\,nm thick $\delta$-MnGa films were grown on \emph{p-i-n} Ga$_{0.9}$Al$_{0.1}$As LEDs with a 10\,nm thick GaAs quantum well (QW) in the intrinsic region. For structural and processing details, see Refs.{} \onlinecite{Adelmann} and \onlinecite{Dong}. Electroluminescence measurements were carried out with a bias voltage applied between the \emph{p}-type substrate and the $\delta$-MnGa contact, so that spin-polarized electrons tunnel from $\delta$-MnGa through the reverse biased Schottky barrier and into the semiconductor heterostructure. The EL was collected along the growth direction, which was parallel to the applied magnetic field. At low temperature, the EL was dominated by recombination of heavy-hole excitons in the GaAs QW. The degree of circular polarization of the EL, $P_{EL} = \left( I^+ - I^- \right) / \left( I^+ + I^- \right)$, was then calculated from the integrated intensities for right ($I^+$) and left ($I^-$) circularly polarized light. The steady-state non-equilibrium electron spin polarization injected into the QW is then $P = P_{EL}$ \cite{DyakonovOO}. The experimental signal $P_{EL}$ also contains a background due to magneto-absorption in the semitransparent $\delta$-MnGa contact. A magneto-absorption measurement on a 7.5\,nm thick film shows that this background is equal to 0.5\% [solid line in Fig.~2(a)].

Figure 2(a) shows $P_{EL}$ as a function of an applied out-of-plane magnetic field at different temperatures, as indicated. A nearly square hysteresis loop is observed at all temperatures, and $P_{EL}$ closely traces the out-of-plane magnetization $M_z$ of $\delta$-MnGa measured by SQUID magnetometry at 10\,K [dashed line in Fig.~2(a)] \cite{Remark}. The observed hysteresis in both $P_{EL}$ and $M_z$ are due to the perpendicular magnetic anisotropy of $\delta$-MnGa, which is large enough to overcome the shape anisotropy of the thin film. We have also observed square hysteresis loops and an EL polarization signal at remanence in $\delta$-MnGa spin LEDs with a bulk detector in place of the QW. The maximum $P_{EL}$ observed in  the $\delta$-MnGa QW structures is approximately 5\% at 2\,K as opposed to almost 30\% in comparable heterostructures with Fe contacts \cite{Adelmann}. Although the spin polarization of bulk $\delta$-MnGa is not known, the lower efficiency may also be due to a partially reacted layer at the $\delta$-MnGa/Ga$_{0.9}$Al$_{0.1}$As interface.

\begin{figure}[tb]
\includegraphics[width=8cm,clip]{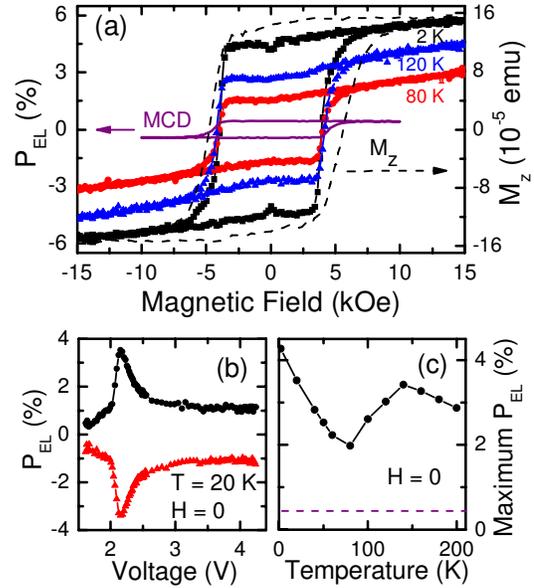}
\caption{\small \textbf{(a)} Magnetic field dependence of the circular polarization of the emitted electroluminescence, $P_{EL}$, from the $\delta$-MnGa spin LED at different temperatures, as indicated. The magnetic field dependence of the out-of-plane magnetization ($M_z$, dashed line) at 10\,K and the magnetoabsorption of the $\delta$-MnGa film at 20\,K (solid line) are also shown. \textbf{(b)} $P_{EL}$ as a function of the bias voltage at 20\,K in remanence after saturating the $\delta$-MnGa layer at $+50$\,kOe (circles) and $-50$\,kOe (triangles), respectively. \textbf{(c)} The maximum value of $P_{EL}$ in remanence as a function of temperature.}
\end{figure}

In addition to the close correspondence between $P_{EL}$ and the magnetization, a second distinctive feature of spin injection into \emph{p-i-n} spin LED detectors is a strong dependence of $P_{EL}$ on the bias voltage \cite{Adelmann,Motsnyi,Dong}. Figure~2(b) shows the bias dependence of $P_{EL}$ in remanence after saturating the magnetization at $\pm 50$\,kOe. The electron spin polarizations measured in the two different remanent states are exactly opposite at all bias voltages. The bias dependence of $P_{EL}$ is qualitatively similar to what has been observed for spin injection from Fe into (Al,Ga)As spin LEDs of similar design \cite{Adelmann}. The similar bias dependence of $P_{EL}$ using Fe and $\delta$-MnGa suggests that the bias dependence is due predominantly to a variation of the electron and spin lifetimes in the semiconductor QW heterostructure.

Figure 2(c) shows the temperature dependence of the maximum $P_{EL}$ observed at each temperature up to 200\,K. The overall temperature dependence, including the minimum at approximately 70\,K and a maximum at 150\,K, is very similar to that found for Fe \cite{Adelmann} and Fe$_x$Co$_{1-x}$ \cite{Jiang} injection contacts. The decrease in $P_{EL}$ between 10 and 70\,K reflects the temperature dependence of the carrier recombination time in the QW. The increase in the signal above 70\,K is more difficult to interpret but has recently been correlated with a rapid increase in the non-radiative recombination rate \cite{Salis}. The decrease at temperatures above 150\,K cannot be described in terms of carrier or spin dynamics alone and is likely due to a decrease in the spin injection efficiency as a consequence of a crossover from tunneling transport to thermionic field-emission as the temperature is increased. 

As is evident from the bias dependence shown in Fig.~2(b), a limitation of the spin LED technique is the fact that the optical polarization $P_{EL}$ depends not only on the spin polarization of the injected carriers $P_0$ but also on relaxation processes in the semiconductor. In the ideal case, negligible spin relaxation occurs during the injection process itself, after which electrons become bound in the quantum well. The steady-state spin polarization $P$ in the quantum well is then determined by the spin relaxation rate and the total recombination rate. The solution of the rate equation determining $P$ in zero magnetic field is well-known from the theory of the optical Hanle effect and is given by $P = P_0/\left(1+\tau_r/\tau_s\right)$ \cite{DyakonovOO}. The sensitivity factor $\left(1+ \tau_r/\tau_s\right)^{-1}$ can be estimated from luminescence experiments under optical pumping \cite{Motsnyi,Strand} or by time-resolved measurements \cite{Salis}. In the case of Fe spin LEDs, some information about the relaxation times can be extracted from the evolution of the polarization $P$ in an oblique magnetic field, as originally demonstrated by Motsnyi \emph{et al.} \cite{Motsnyi}. This is analogous to the optical Hanle effect, although it is difficult to extract quantitative information because the applied magnetic field in the oblique geometry induces rotation of the magnetization of the ferromagnet as well as precession of the spin in the semiconductor.

In the case of perpendicularly magnetized $\delta$-MnGa, we can realize a situation which is identical to that in a traditional optical Hanle effect measurement, in which injected spins are depolarized by a purely transverse magnetic field. Small magnetic fields applied in the plane of the film have a negligible effect on the magnetization of the $\delta$-MnGa film, as can be seen in Fig.~3(a), which shows the perpendicular component $M_z$ of the magnetization (solid curve) as a function of the transverse in-plane magnetic field $H_T$. The orientation of the magnetization remains essentially fixed for $H_T < 5$\,kOe and then slowly rotates into the plane, with a saturation field greater than 30\,kOe. For $H_T < 5$\,kOe, spins are injected along $\hat{z}$, but in the QW they precess about $H_T$. The measured polarization $P_{EL}$ will therefore follow the Hanle form \cite{DyakonovOO}:
\begin{equation}
\label{Hanle}
P = P_0 \frac{1}{1+\tau_r/\tau_s} \cdot \frac{1}{1+\left(\Omega T_s\right)^2},
\end{equation}
where $\Omega = g^* \mu_B B/\hbar$ is the Larmor frequency of electrons in the QW \cite{Malinowski} and $T_s^{-1} = \tau_r^{-1} + \tau_s^{-1}$ is the total spin relaxation rate. $P_{EL}$ is shown as a function of $H_T$ in Fig.~3(a). The measured curve comprises a Lorentzian form with a half-width of approximately 5\,kOe superimposed on a much broader peak with an amplitude of approximately 0.5\%. The sharp peak at small fields is due to the Hanle effect, and the broad background is due to the magneto-absorption in the $\delta$-MnGa, which is gradually suppressed as the magnetization rotates into the plane.

\begin{figure}[tb]
\includegraphics[width=8cm,clip]{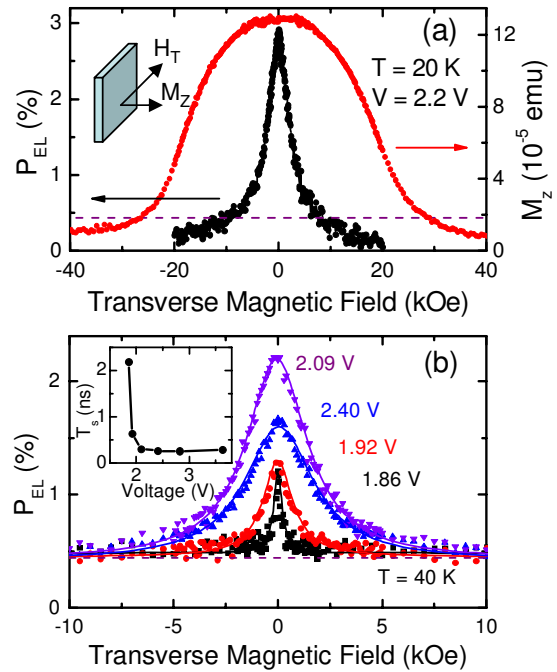}
\caption{\small \textbf{(a)} Electroluminescence polarization $P_{EL}$ (symbols) and out-of-plane magnetization $M_z$ (solid line) of the $\delta$-MnGa film as a function of  the transverse (in-plane) magnetic field.  The inset shows the measurement geometry. \textbf{(b)} $P_{EL}$ is shown as a function of the transverse field for several different bias voltages at a temperature of 40\,K. The solid curves are fits to Eq.~\ref{Hanle}, with a constant offset due to the MCD background (dashed line). The inset shows the bias dependence of the spin lifetime $T_s$ as extracted from the half-widths of these Hanle curves.}
\end{figure}

Data for $P_{EL}$ at small in-plane fields are shown in Fig.~3(b) for several different bias voltages at a temperature of 40\,K. Fits to Eq.~\ref{Hanle} are shown for each bias. A fully quantitative interpretation of the EL measurements requires knowledge of all three parameters $P_0$, $\tau_r$, and $\tau_s$, but only certain combinations of these parameters can be determined from the Hanle curves, as can be seen by inspection of Eq.~\ref{Hanle}. It is clear, however, that the amplitudes and widths of the curves depend markedly on the bias conditions. In particular, the total spin lifetime $T_s$ can be extracted from the measured half-widths, and results for the bias dependence at 40\,K are shown in the inset of Fig.~3(b). The fact that $T_s$ decreases with increasing bias while the amplitude of the EL Hanle curves increases is predominantly due to the decrease in the recombination time $\tau_r$ with increasing voltage. Most importantly, the Hanle curves demonstrate explicitly that the recombination and spin lifetimes are changing dramatically over the bias range of the spin LED measurements.

In summary, we have demonstrated spin injection from $\delta$-Mn$_{0.58}$Ga$_{0.42}$ into (Al,Ga)As heterostructures. The perpendicular magnetic anisotropy of the spin injector allows for the injected polarization to be measured at remanence. Dephasing of the injected spins can be observed in small transverse fields, as indicated by the observation of the Hanle effect.  

We thank X.Y. Dong and M. Nishioka for assistance with the SQUID measurements. This work was supported by the Office of Naval Research, the DARPA SpinS Program, the University of Minnesota MRSEC (DMR 02-12032) and the NSF National Nanotechnology Infrastructure Network through the University of Minnesota Nanofabrication Center.


\begin{thebibliography}{99}
\bibitem{Remark} The linear background that appears in $P_{EL}$ is due to the Zeeman splitting of electron and hole states in the QW and is similar to the background observed in Fe spin LEDs. There is also a local minimum in $P_{EL}$ at zero applied field that becomes more pronounced at lower temperatures. This is likely due to the field dependence of the longitudinal spin lifetime, and, at the lowest temperatures, the presence of hyperfine effects (see Ref. \onlinecite{Strand}).
\end{thebibliography}
\end{document}